\pdfoutput=1
\documentclass{article}
\usepackage{graphicx, multirow}
\usepackage{harvard, authblk, indentfirst, amsmath, hyperref}
\usepackage[margin = 3cm]{geometry}

\begin{document}
\title{%
	Real-time detection and resolution of
	atom bumping in crystallographic models%
}
\author{%
	Yu Liu \texttt{<caspervector@gmail.com>}\\
	National Laboratory for Superconductivity,
	Institute of Physics, Chinese Academy of Sciences,
	Beijing 100190, People's Republic of China;
	University of Chinese Academy of Sciences,
	Beijing 100149, People's Republic of China%
}
\date{}
\maketitle

\section*{Synopsis}

Keywords: collision detection; real-time;
sweep and prune; direct space method; anti-bumping.

Presented here is an $O(n \log n)$ algorithm
that detects atom bonding in a unit cell.
As an application of this algorithm, an evaluation function for atom
bumping is proposed, which can be used for real-time elimination
of crystallographic models with unreasonable bond lengths during the
procedure of crystal structure determination in the direct space.

\section*{Abstract}

A basic principle in crystal structure determination
is that there should be proper distances between adjacent atoms.
Therefore, detection of atom bumping is of fundamental significance
in structure determination, especially in the direct space method
where crystallographic models are just randomly generated.
Presented in this article is an algorithm that detects atom bonding
in a unit cell based on the sweep and prune algorithm of axis-aligned
bounding boxes (AABBs) and running in $O(n \log n)$ time bound,
where $n$ is the total number of atoms in the unit cell.
This algorithm only needs the positions of individual atoms in the unit cell
and does not require any prior knowledge of existing bonds,
and is thus suitable for modelling of inorganic crystals
where the bonding relations are often unknown \emph{a priori}.
With this algorithm, computation routines requiring bonding information,
\emph{eg.}\ anti-bumping and computation of coordination numbers
and valences, can be performed efficiently.
As an example application, an evaluation function for atom bumping is proposed,
which can be used for real-time elimination of crystallographic models
with unreasonably short bonds during the procedure of global optimisation
in the direct space method.

\section{Introduction}\label{sec:intro}

\begin{figure}[htbp]\centering
\includegraphics[width = 0.4\textwidth]{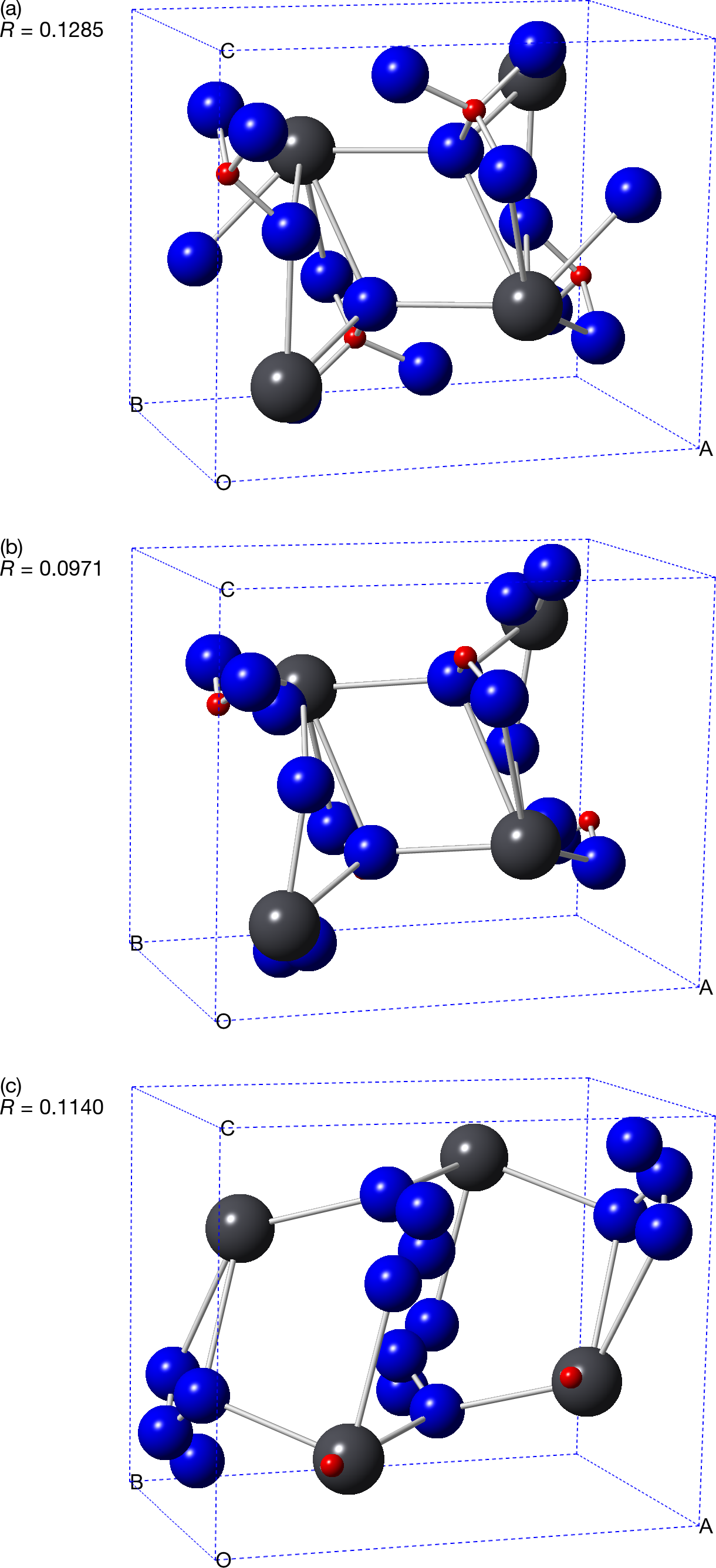}
\caption{Some solutions for the PbSO$_4$ structure, where only (a) is correct}%
\label{fig:pso-bump}
\end{figure}

The direct space method \cite{cerny2007} attempts to determine the structure
of a crystal by finding a coordinate combination for independent atoms
in the unit cell that minimises difference between the observed
diffraction pattern and the pattern computed from the coordinates.
As with the common practice in optimisation algorithms,
coordinates are generated in a randomised fashion,
which often results in crystallographic models
with unreasonably short bonds or, in other words, atom bumping;
sometimes, the pattern computed from an unreasonable crystallographic model
is so similar to the observed pattern, that it becomes a plausible solution
to the above-mentioned optimisation problem.
\emph{Eg.}\ suppose we have obtained the diffraction pattern for PbSO$_4$,
and have somehow known that its space group is $Pnma$,
and that Pb$^{2+}$ and S$^{6+}$ occupy two $4c$ sites while
O$^{2-}$ occupies two $4c$ sites plus one $8d$ site in its structure,
and attempt to determine the exact atomic coordinates with the
direct space method by running the above-mentioned algorithm several times,
the result we usually get would be something like that in Figure
\ref{fig:pso-bump} (see Subsection \ref{ssec:bench} for the data):
most solutions would have several pairs of bumping atoms,
and only the correct solution is free from atom bumping;
however, these solutions' Bragg $R$ factors,
which measure the difference between computed
and observed diffraction patterns, are all similarly small,
so one cannot tell which solution is obviously unreasonable
just by looking at the Bragg $R$ factors.

One way to tackle this problem is to gather a set of solutions,
perform \emph{a posteriori} structure validation \cite{spek2003},
including the bond length check, on them,
and then discard the unreasonable solutions.
However, we know from the example above that
even for crystal structures as simple as PbSO$_4$,
sometimes the solutions are mainly unreasonable ones.
In structure determination of complex structures,
considerable time is spent on the global optimisation procedure,
and consequently it is imperative to automatically eliminate
unreasonable crystallographic models in global optimisation.
Detection of bumping between objects is known
as collision detection in computational geometry,
and what we need here is real-time elimination
of crystallographic models with atom bumping,
which requires real-time collision detection.
Na\"ive pair-wise collision detection requires
collision tests between all $n (n + 1)\,/\,2$
atom pairs in a unit cell containing $n$ atoms,
which is unsatisfactory for large problems, so we need
an efficient crystallographic collision detection algorithm.

The reader might notice that, in the case of PbSO$_4$,
S$^{6+}$ and O$^{2-}$ atoms are grouped into SO$_4^{2-}$ tetrahedra,
so we can simply use the degrees of freedom
of SO$_4^{2-}$ groups as the variables to be optimised,
which would not only eliminate bumping inside each SO$_4^{2-}$ group
by restricting the ranges of internal bond lengths and angles,
but also significantly reduce the complexity
of collision detection between atom groups.
This approach is actually quite successful for structure determination
of molecular crystals \cite{andr1997}, framework crystals \cite{falc1999} and,
to some extent, other crystals \cite{favre2002}.
However, for structure determination of most inorganic crystals,
the bonding relations are usually unknown \emph{a priori}, and sometimes
we have to determine their structures with the direct space method:
high resolution diffraction data can not be obtained for some of them,
especially for nano-materials which can hardly be handled
by single crystal diffraction even with synchrotron radiation,
so reciprocal space methods, \emph{eg.}\ charge flipping \cite{baerl2006},
are of limited use for these crystals.
In these cases, we still need an efficient crystallographic collision
detection algorithm, which is to be presented in this article.

\section{Crystallographic collision detection: the broad phase}\label{sec:broad}

As stated in Section \ref{sec:intro},
na\"ive collision detection requires $O(n^2)$ pairwise tests,
which is one of the biggest obstacles to real-time collision detection.
To solve this problem, the collision detection procedure is
conventionally split into two phases \cite[p.\ 14]{ericson2005}:
the broad phase which prunes pairs of objects
that definitely do not collide,
and the narrow phase which performs exact pairwise tests
between those pairs that may collide.
In this section, we focus on the broad phase of our algorithm.

\subsection{Collision detection with sweep and prune}

A common pattern of the broad phase is to construct
a volume for each object that completely bounds the object,
and perform collision detection with sub-$O(n^2)$ time complexity
on these bounding volumes due to certain geometric properties of the volumes.
A most simple algorithm of this kind is sweep and prune (SAP),
also known as sort and sweep \cite[p.\ 329]{ericson2005},
which uses axis-aligned bounding boxes (AABBs).
We base our own algorithm on SAP,
and briefly introduce the latter in this subsection.

\begin{figure}[htbp]\centering
\includegraphics[width = 0.45\textwidth]{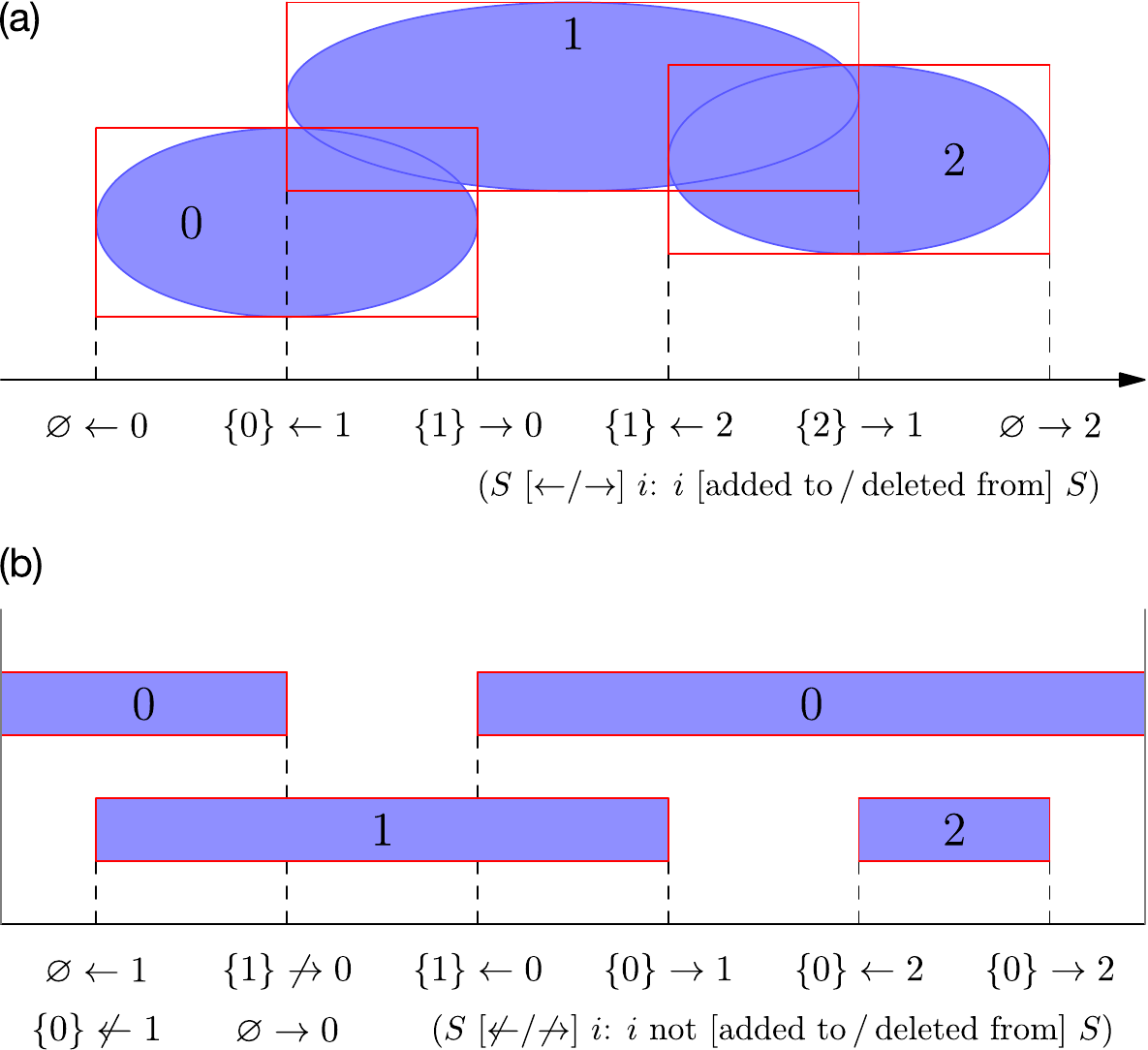}
\caption{Examples of SAP: (a) regular SAP; (b) with the periodic boundary}%
\label{fig:sap-sweep}
\end{figure}

Suppose we want to detect the collision
between $n$ objects (\emph{eg.}\ those in Figure \ref{fig:sap-sweep}(a)).
In SAP, we would first project these objects onto a convenient axis,
and then detect the collision between the projection intervals:
\begin{itemize}
\item The lower and upper bounds of all these projection intervals
	are sorted and inserted into an ordered list $l$ (the sort pass);
	then an empty set $S$ is set up, and the collision between intervals at
	each bound is tracked by sweeping $l$ in ascending order (the sweep pass):
\item When a lower bound is encountered, the collision between
	the corresponding interval and each interval in $S$ is reported,
	and the interval itself is then added to $S$.
	\emph{Eg.}\ in Figure \ref{fig:sap-sweep}(a),
	when the lower bound of interval \#1 is encountered,
	the collision between intervals \#0 and \#1 is reported,
	and then interval \#1 is added to $S$.
\item When an upper bound is encountered,
	the corresponding interval is deleted from $S$.
	\emph{Eg.}\ in Figure \ref{fig:sap-sweep}(a),
	interval \#1 is deleted from $S$ when its upper bound is encountered.
\end{itemize}

In SAP, element insertion and deletion on $S$ are often
part of the output routines and conventionally not counted
in the complexity analysis of the broad phase \cite[p.\ 333]{ericson2005};
nevertheless, it is easy to realise that the time complexity
of these operations depend on the actual number of colliding pairs:
if all objects do collide with every other object,
operations on $S$ would at least require $O(n^2)$ time anyway.
With a decent sorting algorithm,
\emph{eg.}\ mergesort \cite[pp.\ 158--168]{knuth1998},
the sort pass runs in $O(n \log n)$ time for the worst case;
and since the sweep pass obviously runs in $O(n)$ time,
the total time complexity of SAP can be $O(n \log n)$.

\subsection{Implementation notes for sweep and prune}\label{ssec:impl}

\begin{figure}[htbp]\centering
\includegraphics[width = 0.5\textwidth]{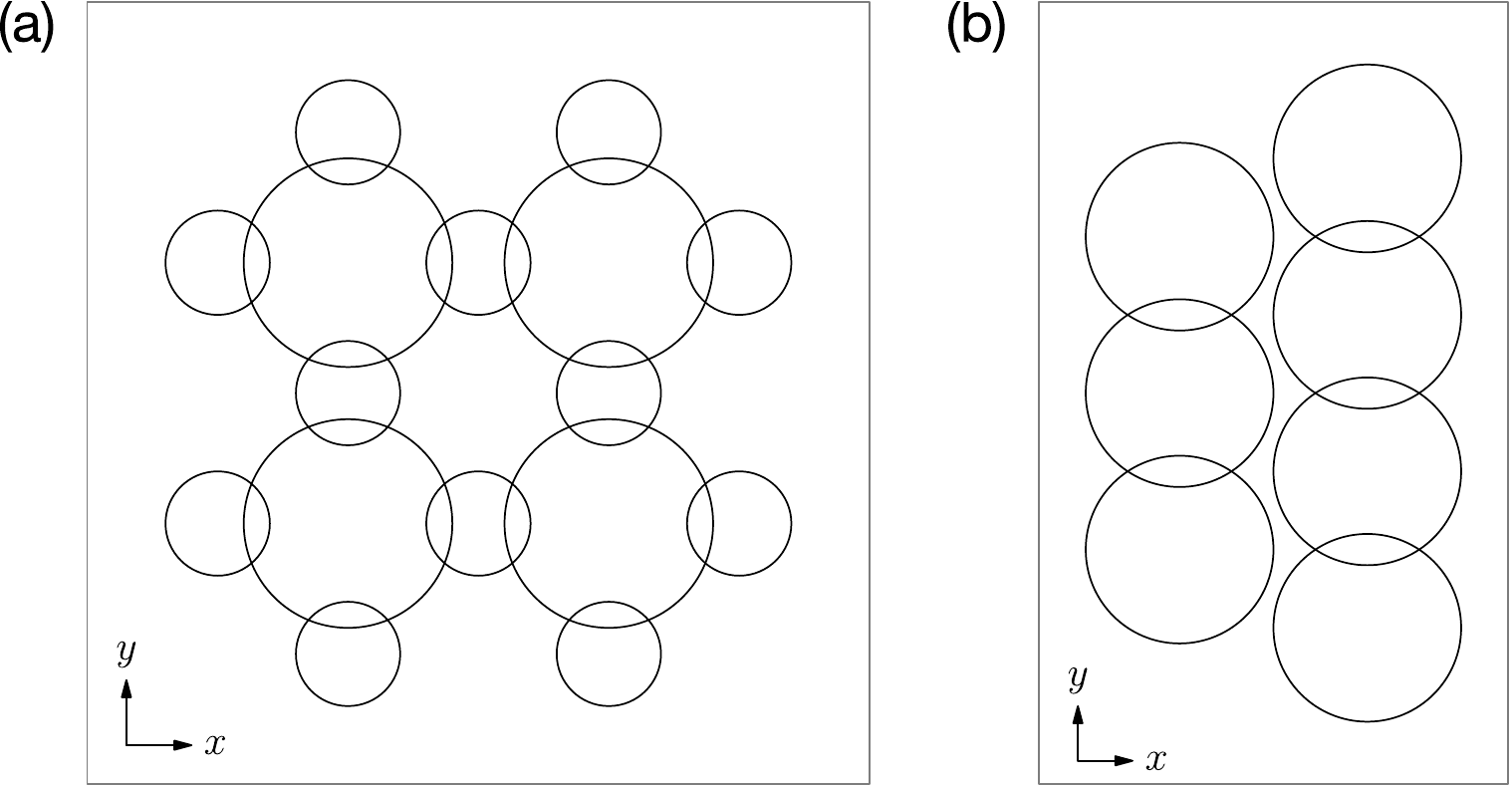}
\caption{%
	Examples of technicalities in the implementation of SAP:
	(a) an example where single-axis SAP would be inefficient;
	(b) an example with objects clustered in the $x$ direction,
		so SAP can be skipped on the $x$ axis%
}\label{fig:mult-axes}
\end{figure}

Some technicalities require attention in the implementation of SAP:
\begin{itemize}
\item For large $n$, non-colliding pairs could be insufficiently pruned
	if only one axis is used (\emph{eg.}\ in Figure \ref{fig:mult-axes}(a),
	using only the $x$ or $y$ axis leads to inefficient collision detection),
	and it is favourable to perform SAP on multiple axes.
	As a side note, when multiple axes are used, the projection procedure
	naturally bounds the objects with boxes, whose boundaries are
	aligned to the used axes (Figure \ref{fig:sap-sweep}(a)),
	hence the notion that SAP uses AABBs.
\item Sometimes objects are clustered in some directions
	(\emph{eg.}\ the $x$ direction in Figure \ref{fig:mult-axes}(b)),
	resulting in inefficient collision detection on corresponding axes,
	and it is thus advisable to skip SAP on them.
\item It can often be assumed that a pair of objects
	do not actually interact if they only collide at their boundaries
	(\emph{eg.}\ two balls that are tangent to each other do not bump),
	so if some lower and upper bounds occur at a same position in $l$,
	the upper bounds can be prioritised over the lower bounds in the sweep pass.
\item Although the quicksort algorithm \cite[pp.\ 113--122]{knuth1998}
	can be $O(n^2)$ for the worst case,
	it is $O(n \log n)$ in average and has a smaller constant factor,
	and is therefore often more suitable for the sort pass.
\end{itemize}

In addition, noticing that the normal bond length between two atoms can be
different from the sum of their radii because of the Coulomb interaction,
here we introduce the \emph{pairwise zoom factor} $p$,
which is the ratio of the normal bond length to the sum of radii.
We also introduce the \emph{collision detection radius},
which is the radius that the projection interval of an atom is computed from;
the collision detection radius $r$ of an atom is its normal radius $r_0$
multiplied by its \emph{atomic zoom factor} $q$.
For correctness of the broad phase, it is necessary to guarantee
that for any atom pair $\{k_0, k_1\}$, we have
\[ p(k_0, k_1) (r_0(k_0) + r_0(k_1)) \leq q(k_0) r_0(k_0) + q(k_1) r_0(k_1). \]

\subsection{Handling non-orthogonality of the unit cell}

\begin{figure}[htbp]\centering
\includegraphics[width = 0.5\textwidth]{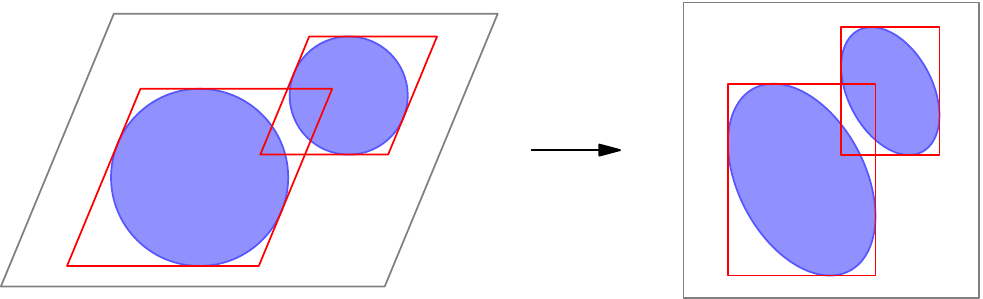}
\caption{%
	Equivalence between using axis-aligned projection
	and using AABBs in fractional coordinates%
}\label{fig:aabb-frac}
\end{figure}

We know the $a$, $b$, $c$ axes of the unit cell may well be non-orthogonal,
so cuboid bounding boxes would be cumbersome to use in the unit cell.
However, the original idea of SAP, \emph{ie.}\ detecting the collision
between projection intervals on some axes, can still be easily be applied,
just with the difference that the projection is not necessarily orthogonal.
Here we use \emph{axis-aligned projection}, which is equivalent to using
non-cuboid AABBs in Cartesian coordinates, which are in turn equivalent to
regular AABBs in fractional coordinates (Figure \ref{fig:aabb-frac}).
Since fractional coordinates are already heavily used in crystal structure
determination, from the figure we know it is natural to use
fractional coordinates in crystallographic collision detection,
so now we only need to solve one problem:
given the Cartesian radius and fractional coordinates of an atom,
how do we compute its AABB in fractional coordinates?

\begin{figure}[htbp]\centering
\includegraphics[width = 0.5\textwidth]{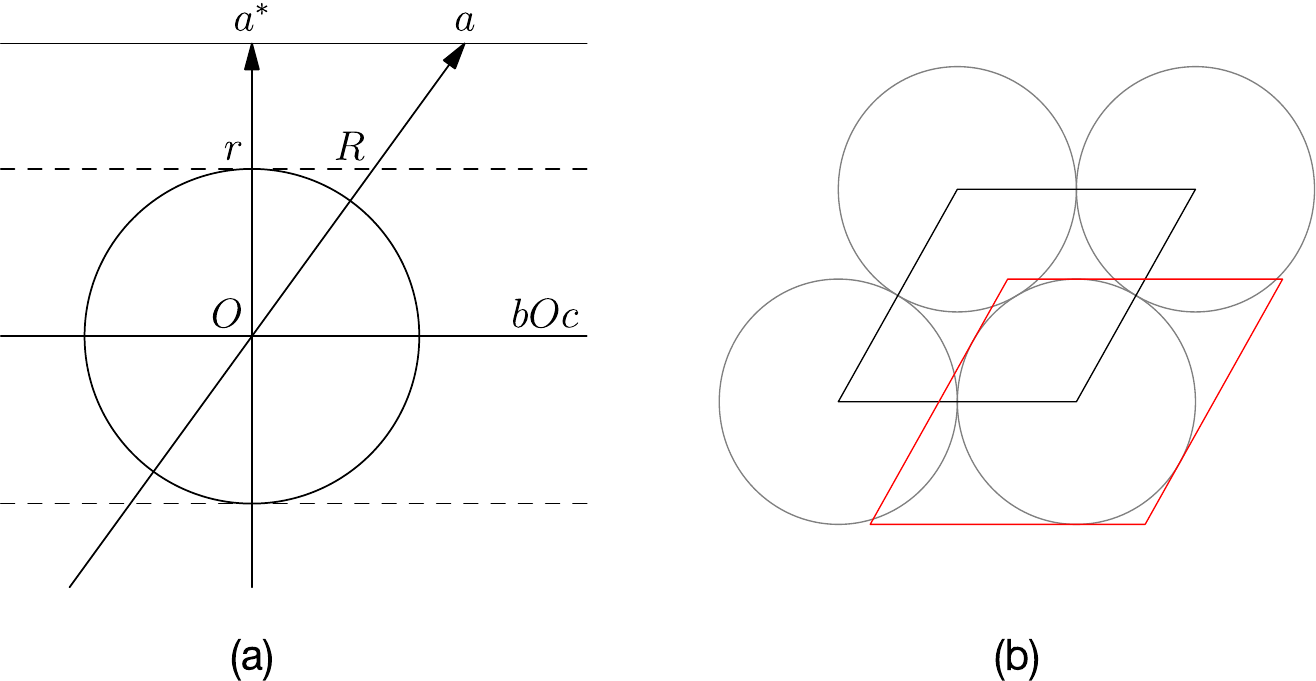}
\caption{%
	(a) Projection of an atom onto the $a$ axis;
	(b) a scenario where the bounding box (light)
		is larger than the unit cell (dark)%
}\label{fig:aabb-width}
\end{figure}

Consider the projection of an atom with
collision detection radius $r$ onto the $a$ axis of a unit cell:
what we want to compute is the half-width $R$ of the projection interval.
When viewed from a direction orthogonal to both the $a^*$ axis
(and therefore parallel to the $bOc$ plane) and the $a$ axis
(Figure \ref{fig:aabb-width}(a)),
according to the properties of similar triangles we know that
\[
	\frac{R}{a} = \frac{r}{d} \Rightarrow
	\frac{R}{ar} = \frac{1}{d} = a^* = \frac{\sin\alpha}{av},
\]
where $d$ is the distance between two adjacent $bOc$ nets, and
\[
	v = V / abc = \sqrt{
		1 - \cos^2 \alpha - \cos^2 \beta - \cos^2 \gamma
		+ 2 \cos\alpha \cos\beta \cos\gamma
	}
\]
is the nondimensionalised volume of the unit cell.
The formulae for $a^*$ and $V$ are courtesy of \citeasnoun{prince2004}.

Since fractional coordinates need to be used finally,
the half-width we actually use should be
\[ \frac{R}{a} = \frac{r}{d} = \frac{r\sin\alpha}{av}; \]
the same goes for the $b$ and $c$ axes.
Note that since atoms tend to cluster in directions with disproportionately
(among $a$, $b$ and $c$) large $R\,/\,ar$ ratios, it is advisable
to skip SAP on corresponding axes (\emph{cf.}\ Subsection \ref{ssec:impl}).

\subsection{Handling the periodic boundary}

SAP tracks the collision between intervals
at each (lower or upper) bound by sweeping $l$, the list of bounds.
The correctness of the sweep pass relies on occurrence
of the lower bound of each interval before the upper bound,
which does not necessarily hold for the unit cell
due to its periodic boundary (Figure \ref{fig:sap-sweep}(b)).
This results in upper halves of the cross-boundary intervals
being neglected in the sweep pass:
\emph{eg.}\ in Figure \ref{fig:sap-sweep}(b), the collision between
intervals \#0 and \#1 is neglected.

Nevertheless, the issue of cross-boundary objects is usually easy to resolve:
just additionally test the collision involving these cross-boundary objects.
In SAP with periodic boundary, at the end of the sweep pass, $S$, the set
of intervals at the current position, contains all cross-boundary intervals,
so we can modify the original algorithm to:
\begin{itemize}
\item Perform the original algorithm, but take care
	when an upper bound is encountered:
	if the corresponding lower bound have not be encountered yet,
	the corresponding interval would be absent from $S$,
	therefore it would be meaningless to delete the interval from $S$.
	\emph{Eg.}\ in Figure \ref{fig:sap-sweep}(b),
	when the upper bound of interval \#0 is encountered,
	we cannot delete the interval from $S$.
\item Again sweep $l$ in ascending order, but never add an interval to $S$,
	because collision between cross-boundary intervals have already
	been reported at the end of the original sweep pass,
	so we only care about the collision between \{cross-boundary interval,
	regular interval\} pairs.
	\emph{Eg.}\ in Figure \ref{fig:sap-sweep}(b),
	when the lower bound of interval \#1 is encountered again,
	the interval is not added to $S$.
\item Terminate when $S$ becomes empty.
	\emph{Eg.}\ in Figure \ref{fig:sap-sweep}(b), the algorithm terminates
	when the upper bound of interval \#0 is encountered again.
\end{itemize}

In worst scenario, the additional sweep pass requires $O(n)$ time,
so the modified algorithm can still run in $O(n \log n)$ time.
However, some caveats apply to the modified algorithm:
\begin{itemize}
\item In the additional sweep pass,
	duplicate collision pairs might be reported when
	an interval bounds the complement of a cross-boundary interval:
	\emph{eg.}\ in Figure \ref{fig:sap-sweep}(b),
	the collision between intervals \#0 and \#1 is reported twice.
	These duplicates need to be handled properly.
\item This algorithm requires the width of an interval
	to be always shorter than the width of the axis,
	so it can tell whether an interval cross the boundary by examining
	whether the upper bound is less than the lower bound.
	This condition can be violated under some extreme conditions,
	\emph{eg.}\ in Figure \ref{fig:aabb-width}(b).
	To solve this problem, we can hard-limit the half-width
	(in fraction coordinate) of an interval to $0.5 - \epsilon$,
	where $\epsilon$ is a very small positive constant.
\end{itemize}

\subsection{Exploiting equivalent positions}

\begin{figure}[htbp]\centering
\includegraphics[width = 0.6\textwidth]{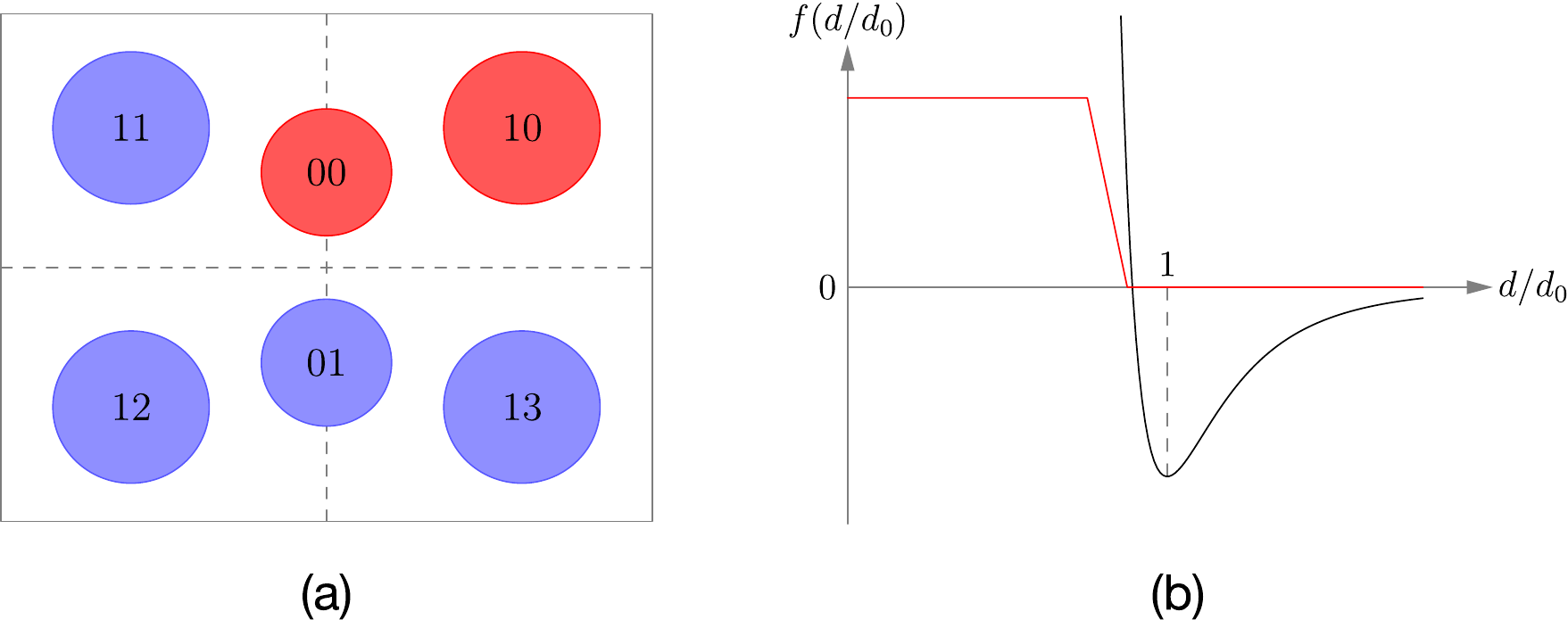}
\caption{%
	(a) An example of equivalent positions;
	(b) the actually used two-body potential (light)
		in comparison with the Lennard-Jones potential (dark)%
}\label{fig:eq-eval}
\end{figure}

Let the independent atoms in a unit cell be numbered 0, 1,
$\cdots, m - 1$ and labeled by index $i$, with the equivalent atoms
of independent atom $i$ labeled $ij$, with $j =$ 0, 1, $\cdots, n_i - 1$:
\emph{eg.}\ in Figure \ref{fig:eq-eval}(a), the second equivalent atom
of the first independent atom is numbered 01 ($i = 0, j = 1$).
Consider a two-body function
\[ c(i_0j_0, i_1j_1) = c(i_1j_1, i_0j_0)\ (i_0j_0 \neq i_1j_1) \]
satisfying the equivalent position symmetry:
for any symmetry operation $T$ compatible with the unit cell, we have
\[ c(i_0j_0, i_1j_1) = c(T(i_0j_0), T(i_1j_1)). \]
The Euclidean distance function, as an example, satisfies this condition.

For any atom $i_0j_0$, there must be a symmetric operation $T_{i_0j_0}$
transforming it into atom $i_00$ since they are equivalent atoms, so we have
\[
	\sum_{i_1j_1 \neq i_0j_0}\hspace*{-1.5ex} c(i_0j_0, i_1j_1)
	= \hspace*{-1.5ex}\sum_{i_1j_1 \neq i_0j_0}\hspace*{-1.5ex}
		c(i_00, T_{i_0j_0}(i_1j_1))
	= \hspace*{-1.1ex}\sum_{i_1j_1 \neq i_00}\hspace*{-1.1ex} c(i_00, i_1j_1),
\]
which leads to
\[
	\sum_{\{i_0j_0, i_1j_1\}}\hspace*{-1.9ex} c(i_0j_0, i_1j_1)
	= \frac12\sum_{i_0} n_{i_0}
		\hspace*{-1.1ex}\sum_{i_1j_1 \neq i_00}\hspace*{-1.1ex} c(i_00, i_1j_1)
	= \hspace*{-1.9ex}\sum_{\{i_0j_0, i_1j_1\}}\hspace*{-1.9ex}
		\delta_{j_0} n_{i_0} c(i_0j_0, i_1j_1),
\]
where
\[ \delta_j = 1\ \text{if}\ j = 0\ \text{else}\ 0 \]
is the Kronecker delta symbol.

Therefore we also have
\[
	\sum_{\{i_0j_0, i_1j_1\}}\hspace*{-1.9ex} c(i_0j_0, i_1j_1)
	= \hspace*{-1.9ex}\sum_{\{i_0j_0, i_1j_1\}}\hspace*{-1.9ex} c(i_1j_1, i_0j_0)
	= \hspace*{-1.9ex}\sum_{\{i_0j_0, i_1j_1\}}\hspace*{-1.9ex}
		\delta_{j_1} n_{i_1} c(i_0j_0, i_1j_1).
\]
Averaging the two equations above results in the finally used equation:
\begin{equation}\label{eq:eq-pos}
	C = \hspace*{-1.9ex}\sum_{\{i_0j_0, i_1j_1\}}\hspace*{-1.9ex} c(i_0j_0, i_1j_1)
	= {\textstyle\frac12} \hspace*{-1.9ex}\sum_{\{i_0j_0, i_1j_1\}}\hspace*{-1.9ex}
		(\delta_{j_0} n_{i_0} + \delta_{j_1} n_{i_1}) c(i_0j_0, i_1j_1).
\end{equation}

Note that if we only set $j$ of an atom to 0
when it lies in a chosen asymmetric unit,
the $(\delta_{j_0} n_{i_0} + \delta_{j_1} n_{i_1})$ term
would be always zero for $\{i_0j_0, i_1j_1\}$ pairs outside of the unit:
\emph{eg.}\ in Figure \ref{fig:eq-eval}(a), we get
\[
	C = \frac12 \Big((2 + 4) c(00, 10) + 2c(00, \{01, 11, 12, 13\})
		+ 4c(10, \{01, 11, 12, 13\}) \Big).
\]
The significance of this is that computation of $c(i_0j_0, i_1j_1)$ for atom
pairs outside of the asymmetric unit can be skipped even without SAP, leading to
a significant decrease in the number of pairwise tests from $n (n - 1)\,/\,2$ to
$mn - m (m - 1)\,/\,2$ (and consequently, from $O(n^2)$ to $O(mn)$).

\section{Crystallographic collision detection: the narrow phase}

With the broad phase algorithm presented in Section \ref{sec:broad},
we can efficiently prune atom pairs that either definitely do not collide
or can be skipped due to the equivalent position symmetry.
In this section, we discuss the narrow phase of our algorithm,
propose an evaluation function for atom bumping as an example application,
and finally evaluate the effectiveness and efficiency of our algorithm.

\subsection{Bond length computation and the closest vector problem}%
\label{ssec:narrow}

\begin{figure}[htbp]\centering
\includegraphics[width = 0.5\textwidth]{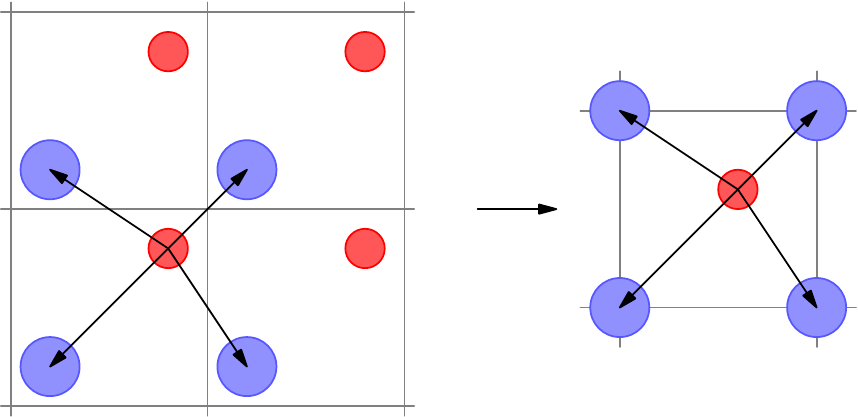}
\caption{Equivalence between bond length computation and CVP}%
\label{fig:cryst-cvp}
\end{figure}

When talking about length of the bond between two atoms in a unit cell,
we usually mean the shortest distance between two lattices
constructed by periodic translation of the two atoms,
which is equal to the shortest distance between the interatomic displacement
vector and the lattice constructed from the cell origin
due to translational symmetry (Figure \ref{fig:cryst-cvp}).

\begin{figure}[htbp]\centering
\includegraphics[width = 0.4\textwidth]{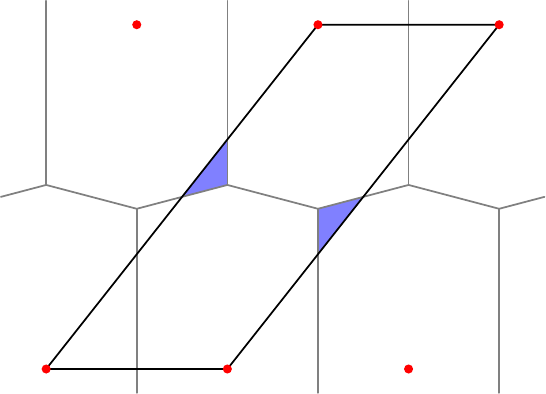}
\caption{%
	A badly chosen unit cell, where the closest lattice point to any point
	in the shaded area lies outside of the given cell; the plane is
	partitioned according to the closest lattice point to each point%
}\label{fig:bad-cell}
\end{figure}

This problem is obviously close-related to the closest vector problem
(CVP) \cite{micc2002}: given a lattice and a vector,
find a closest lattice point to the given vector.
CVP is interesting in cryptography, in large part because
its computational complexity grows rapidly with regard to dimension
of the used vector space, even with quantum computers \cite{bern2009};
the problem is easier to solve if a basis,
that consists of short and nearly orthogonal basis vectors,
can be found for the given lattice.
In crystallography, since the involved dimension does not often exceed 3,
the computational complexity of CVP is not a big problem,
but we still need to perform lattice reduction:
in our algorithm, we assume that a suitable cell choice is used,
so that the closest lattice point to a vector can never be
outside of a cell that bounds the vector;
a counterexample for this is shown in Figure \ref{fig:bad-cell}.

With the assumption above, we can compute the bond length between two atoms
by computing the displacement vector between the atoms and finding the
shortest distance between the vector and the vertices of the bounding cell.
This is clearly faster than the common approach \cite{attf1999,grud2010}
of finding the shortest distance between one atom and replicas of
the other atom from the bounding cell and all of its adjacent cells:
\emph{eg.}\ in the 3-dimensional space, we need to compute vector lengths
27 times in the common approach, and only 8 times in our approach.

We also note that, for interatomic displacements with $l^1$ norms
\[ |\Delta x| + |\Delta y| + |\Delta z| < 1 / 2 \]
in fractional coordinates, an author \cite{krish2015}
directly treats length of the displacement vector as the bond length,
however we did not find a proof for its correctness.
This approach is also used in SHELX \cite{sheld2010},
but without an explicit restriction on the $l^1$ norm.

\subsection{An evaluation function for atom bumping}\label{ssec:eval}

As mentioned in Section \ref{sec:intro}, the direct space method
abstracts crystal structure determination as the optimisation problem
of finding a best coordinate combination for independent atoms
\[ \mathbf{x} = (x_1, y_1, z_1, x_2, y_2, z_2, \ldots, x_m, y_m, z_m) \]
that minimises difference between the computed and the observed
diffraction patterns, which is usually measured by the Bragg $R$ factor.
This approach does not consider the atom bumping problem,
so we might get chemically unreasonable
crystallographic models with small $R$ factors;
on the other hand, if we somehow take atom bumping into account
during the procedure of global optimisation, we would
be able to automatically eliminate these unreasonable models.

One way to account for atom bumping is to modify
the constraints for the optimisation problem,
so we always generate crystallographic models without atom bumping;
however, this greatly complicates the optimisation procedure,
most importantly because the set of acceptable models would
be fragmented into many disjoint chunks in the domain of $\mathbf{x}$,
conflicting with the pattern of random continuous moves
in global optimisation and resulting in inefficient optimisation.
Instead, here we choose to modify the objective function:
since a small $R$ factor does not guarantee the corresponding
crystallographic model to be free from atom bumping,
we can explicitly combine it with an evaluation function that
penalises atom bumping, so that solutions without atom bumping are
more likely to be produced by global optimisation.

In optimisation algorithms, the objective function to be minimised
is often considered as the energy of some physical system,
so to optimise the function is to minimise the energy of the system.
Following this line of thought,
our evaluation function for atom bumping is based on a two-body potential
\[ C = \sum_{\{k_0, k_1\}} c(k_0, k_1), \]
where $k$ is the index of an atom in the unit cell.
The pairwise potential function $c$ is defined as
\[ c(k_0, k_1) = f(d(k_0, k_1) \big/ d_0(k_0, k_1)), \]
where $d$ is the bond length, $d_0$ is the normal bond length,
and $f$ is a piecewise linear function with control nodes
\[ f(0) = f(0.75) = 1, f(0.875) = f(+\infty) = 0. \]
The shape of $f$ is designed to mimic the Lennard-Jones potential
but with a less steep slope, as shown in Figure \ref{fig:eq-eval}(b).
Since we only begin to consider two atoms as bumping when the bond length
between them is 0.875 times the normal bond length,
the collision detection radius of an atom needs to be computed as
\[ r(k) = 0.875 q(k) r_0(k). \]

It is obvious that the pairwise potential presented here
satisfies the equivalent position symmetry, so we can compute
the value of $C$ using equation (\ref{eq:eq-pos}).
For large $n$, the average value of $C$ for a random model
would also be large due to more bumping atom pairs.
To address this, we define our evaluation function as
\[ B = \min(C\,/\,n, 1); \]
the function is hard-limited to 1 because when $C = n$,
there would be two atoms in full bumping with every atom in average,
which should be considered very unreasonable in our opinion.

Since $B \in [0, 1]$, it would be nice
if an upper bound $\nu$ can be obtained for the Bragg $R$ factor,
so we can define the objective function as
\[ E = \mu B + (1 - \mu) R\,/\,\nu\ (\mu \in [0, 1]), \]
which makes $E \in [0, 1]$ for any value of $\mu$, the combination factor.
Provided that normalised intensities are used for both patterns, we have
\[
	R = \sum_i |I_{\text{obs}, i} - I_{\text{calc}, i}|
		\Big/ \sum_i I_{\text{obs}, i}
	= \sum_i |I_{\text{obs}, i} - I_{\text{calc}, i}| = 2D,
\]
where $D \in [0, 1]$ is the total variation distance \cite{levin2008} between
the two normalised diffraction patterns viewed as finite probability measures.
Therefore we have $\nu = 2$, and get the formula for the objective function:
\[ E = \mu B + (1 - \mu) D = \mu B + (1 - \mu) R\,/\,2. \]

\subsection{Evaluation of the presented algorithm}\label{ssec:bench}

A part of the code for this article have been
open-sourced as part of our ongoing crystallographic software project,
and can be obtained at \url{https://gitlab.com/CasperVector/decryst/}.
Test code and data for this subsection,
as well as data for Figure \ref{fig:pso-bump},
can be obtained from the supplementary materials.

\begin{table}[htbp]\centering
\caption{Parameters and results for evaluation of our algorithm}%
\label{tbl:alg-eval}
\begin{tabular}{ll}\hline
Cell parameters &
$Pnma; (a, b, c)$ (\AA): (8.4720, 5.3973, 6.9549) \\
\multirow{2}{20ex}{Elements (radii (\AA), zoom factors): site occupations} &
Pb$^{2+}$ (1.33, 1): $1 \times 4c$ \\&
S$^{6+}$ (0.43, 2.8): $1 \times 4c$ \\&
O$^{2-}$ (1.26, 1): $2 \times 4c + 1 \times 8d$ \\
Pairwise zoom factors$^*$ &
S$^{6+}$--Pb$^{2+}$: 1.4; S$^{6+}$--S$^{6+}$: 2.8;
S$^{6+}$--O$^{2-}$: 0.9 \\
Counts of pairwise tests &
$n_\text{orig}$: 276; $n_\text{max}$: 105;
$n_\text{real}$: 17(5); $n_\text{eff}$: 8(3) \\
Time per model (ns) & \multirow{2}{50ex}{%
	$t_\text{void}$: 721(49);
	$t_\text{bn}$: 1032(70); $t_\text{bN}$: 7090(388);
	$t_\text{Bn}$: 25468(558); $t_\text{BN}$: 26662(730)%
} \\\\\hline
\multicolumn{2}{l}{* Pairwise zoom factors default to 1.}
\end{tabular}
\end{table}

To test the effectiveness of our broad phase algorithm,
we ran it with 10000 random crystallographic models generated for PbSO$_4$
with the control parameters specified in Table \ref{tbl:alg-eval}.
We computed the maximum numbers of pairwise tests for each model
with and without using equation (\ref{eq:eq-pos})
($n_\text{max}$ and $n_\text{orig}$, respectively),
and collected the averages and standard deviations of number
of actual pairwise tests ($n_\text{real}$) and number
of pairwise tests that returned affirmative results ($n_\text{eff}$)
for each model, as shown in Table \ref{tbl:alg-eval}.
From the table we can see that our broad phase algorithm
effectively pruned the non-colliding atom pairs for PbSO$_4$.

To test the efficiency of our algorithm,
we ran a test routine in different conditions,
each condition with 250 groups of crystallographic models,
each group consisting of 250 random models generated for PbSO$_4$
with the control parameters specified in Table \ref{tbl:alg-eval}:
\begin{itemize}
\item ($t_\text{void}$)
	no collision detection at all, only random model generation.
\item ($t_\text{bn}$)
	no SAP, only using equation (\ref{eq:eq-pos});
	the narrow phase is a do-nothing routine.
\item ($t_\text{bN}$)
	no SAP, only using equation (\ref{eq:eq-pos});
	the narrow phase is performed as usual.
\item ($t_\text{Bn}$) SAP is performed,
	but the narrow phase is a do-nothing routine.
\item ($t_\text{BN}$) full collision detection.
\end{itemize}
We collected the averages and standard deviations of time
consumed for each group divided by the number of models in each group
on an Intel i7-3720QM processor, as shown in Table \ref{tbl:alg-eval}.
From the table we can see that while SAP is $O(n \log n)$,
its constant factor is quite large, reminding us of the comparison
between mergesort and quicksort in Subsection \ref{ssec:impl};
thus it is better to use only equation (\ref{eq:eq-pos})
in the broad phase for small unit cells.

\begin{figure}[htbp]\centering
\includegraphics[width = 0.45\textwidth]{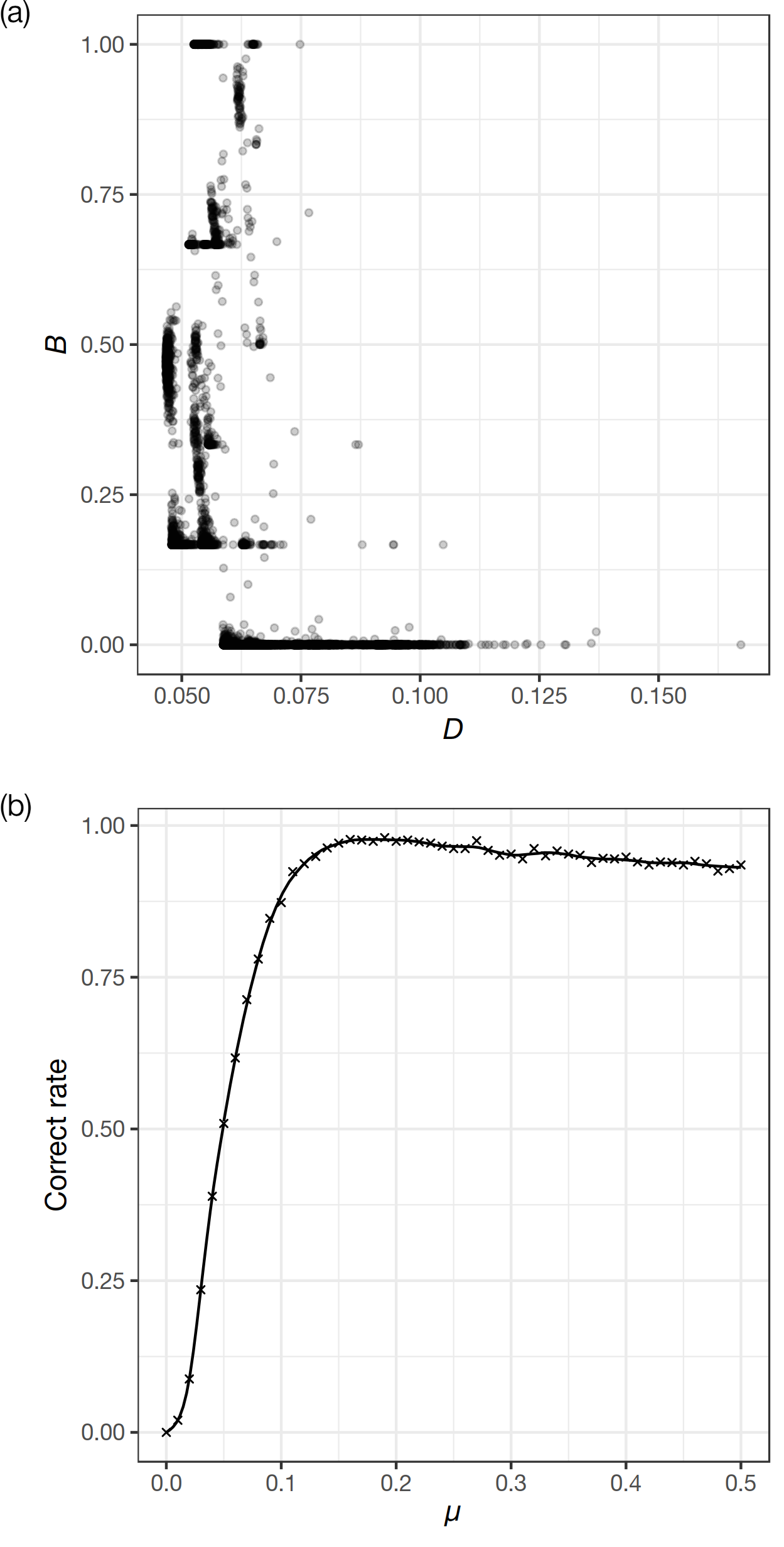}
\caption{%
	(a) $(D, B)$ distribution of 1000 solutions for each value of
		$\mu =$ 0, 0.01, 0.02, $\cdots$, 0.5;
	(b) experimental correct rates of solutions for different $\mu$ values%
}\label{fig:db-distrib}
\end{figure}

\begin{figure}[htbp]\centering
\includegraphics[width = 0.4\textwidth]{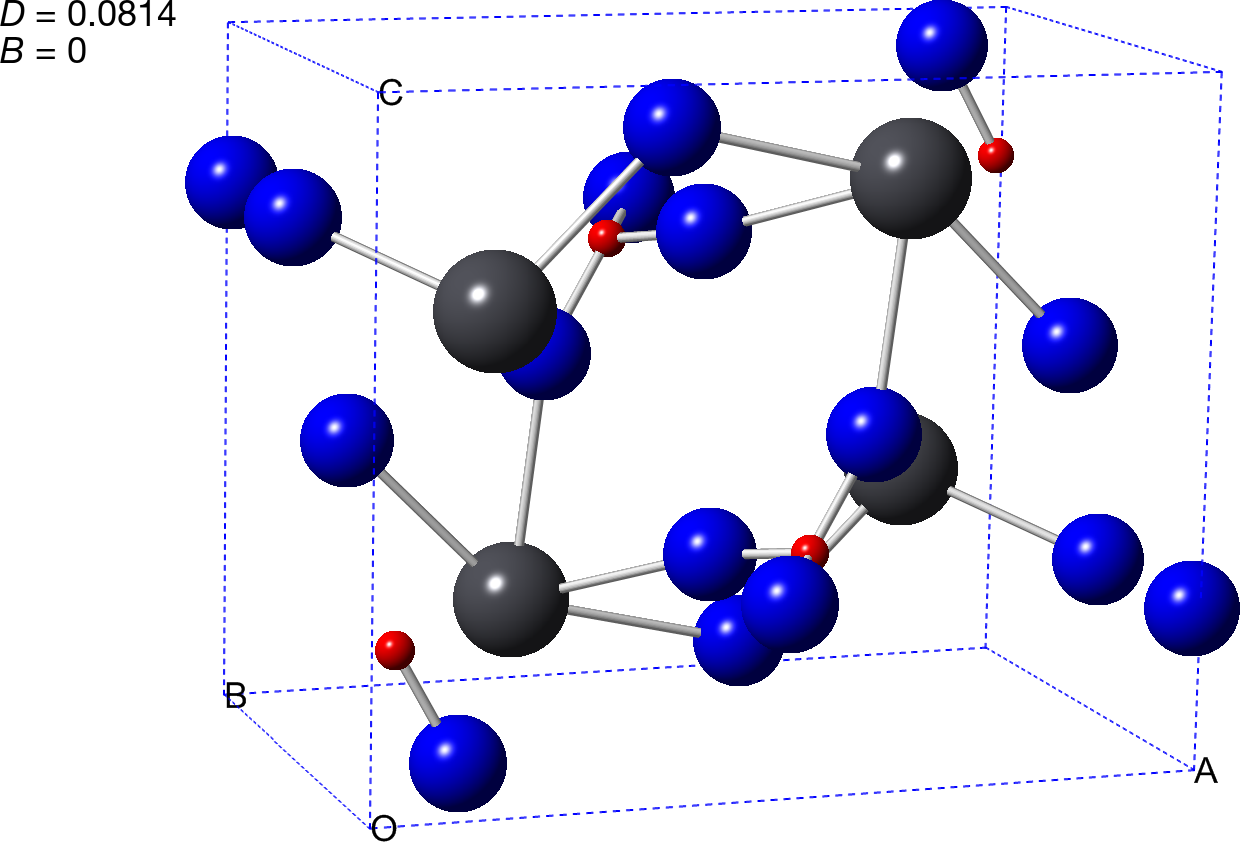}
\caption{An incorrect solution with nearly reasonable $(D, B)$ values}%
\label{fig:verify-fail}
\end{figure}

We also tested the effectiveness of the evaluation function
proposed in this article by running global optimisation for PbSO$_4$
with the control parameters specified in Table \ref{tbl:alg-eval},
using $E$ from Subsection \ref{ssec:eval} as the objective function.
We generated 1000 random solutions for each value of $\mu =$
0, 0.01, 0.02, $\cdots$, 0.5 (we confine ourselves to this range
since our ultimate goal is crystal structure determination after all),
and plotted the $(D, B)$ distribution in Figure \ref{fig:db-distrib}(a).
Since both $D$ and $B$ are continuous with regard to atomic coordinates
and invariant under symmetric transforms,
noticing that the correct solution has $D = R\,/\,2 = 0.0642$ and $B = 0$,
from the figure we consider it appropriate to assume all correct
(up to symmetric transforms and minor fluctuations in atomic coordinates)
solutions to have $D < 0.12$ and $B < 0.05$.
To check this, we generated 100 random solutions
with $D < 0.12$ and $B < 0.05$ using random $\mu \in [0, 0.5]$,
and manually verified each solution;
we found 99 of them to be correct, and the only incorrect solution
(Figure \ref{fig:verify-fail}) had $D = 0.0814$ (the greatest
in these solutions, with the second greatest being 0.0705) and $B = 0$.
According to this, we assume that a solution is correct if and only if it has
$D < 0.075$ and $B < 0.05$; we then collected the ratio of correct solutions
in the 1000 random solutions generated previously for each value of $\mu$,
as shown in Figure \ref{fig:db-distrib}(b).
From the figure we conclude that our evaluation function effectively
eliminated unreasonable crystallographic models for PbSO$_4$.

\section{Discussion and conclusion}

\subsection{Discussion}

Equation (\ref{eq:eq-pos}) does not fully exploit
the symmetry of special positions: \emph{eg.}\ in Figure \ref{fig:eq-eval}(a),
since $c(00, 11) = c(00, 10)$ and $c(00, 13) = c(00, 12)$,
we can actually reduce $C$ to
\[ 4c(00, 10) + c(00, 01) + 2c(00, 12) + 2c(10, \{01, 11, 12, 13\}). \]
If this kind of reduction can be done automatically,
we would be able to further simplify the computation of $C$;
however, it would probably require a much more elaborate
formula for pair multiplicities than the simple
$(\delta_{j_0} n_{i_0} + \delta_{j_1} n_{i_1})\,/\,2$
in equation (\ref{eq:eq-pos}).

As mentioned in Subsection \ref{ssec:narrow}, we assume
in our narrow phase algorithm that a suitable cell is chosen so that
the closest lattice point to a vector cannot be outside of a bounding cell.
This requires that at least one such cell choice exists for every lattice,
which is unproved to our knowledge, although we consider it promising;
we hope theorists can prove or disprove this condition and,
if it is correct, provide an algorithm to construct suitable cell choices.
On the other hand, if this assumption is proved incorrect,
our narrow phase would need to search more neighbour cells for the
closest lattice point to a displacement vector,
but our broad phase would need no change.

The convenient algorithm in Subsection \ref{ssec:narrow}
for computing bond lengths with $l^1$ norms less than
$1\,/\,2$ seems, though unproved, promising to us:
if it is proved correct, we would be able to dramatically
reduce the complexity for computing small bond lengths.
However we note that additional restrictions, like some restrictions on the
cell choice, would be necessary; otherwise incorrect results can be produced,
\emph{eg.}\ for the unit cell in Figure \ref{fig:bad-cell}.
We also note that if no broad phase is performed on the atoms
except for the use of equation (\ref{eq:eq-pos}),
atom pairs with long distances would not be pruned,
so the performance enhancement from this algorithm would be limited.

More elaborate two-body potentials, like those used by \citeasnoun{bushm2012},
or even attractive potentials, could also be used in our evaluation function.
We modeled $c$ from the repulsive section of the Lennard-Jones potential, since
we think our goal is anti-bumping, not precise modeling of atomic interactions.
Nevertheless, we found $c$ to be, while deceptively simple,
surprisingly effective at least for some small unit cells like PbSO$_4$.

\subsection{Conclusion}

We presented an $O(n \log n)$ algorithm for
crystallographic collision detection;
for small unit cells, we also presented an algorithm
that significantly outperforms the na\"ive algorithm.
Based on our algorithm, we proposed an evaluation function
for atom bumping, which can be used for real-time elimination
of unreasonable crystallographic models.

\subsection*{Acknowledgements}

The author would like to thank Cheng Dong for bringing up the issue of
crystallographic collision detection, thank the American Mineralogist
Crystal Structure Database \cite{downs2003} for kindly providing
crystallographic data in the spirit of open access, and thank
the authors of the GNU parallel software package \cite{tange2011} for
providing a convenient parallelism toolkit for the Unix command line.
This project was supported by the National Natural Science
Foundation of China under grant No.\ 21271183.

\bibliographystyle{agsm}
\bibliography{art1}
\end{document}